\documentstyle{article}

\newcommand{\be}{\begin{equation}}
\newcommand{\ee}{\end{equation}}
\newcommand{\mb}{\mbox}
\newcommand{\fr}{\frac}

\newcommand{\p}{\partial}
\newcommand{\pr}{\prime}

\begin{document}

\title{On the large-scale structure of the inflationary universe}
\author{ Jelle P. Boersma \\ ~ \\
Department of Applied Mathematics, University of Cape Town \\
Rondebosch 7701, Cape Town, South Africa}
\maketitle
\begin{abstract}
We study the evolution of quantum fluctuations
of a scalar field which is coupled to the 
geometry, 
in an
exponentially 
expanding
universe.
We derive an expression for the spectrum of intrinsic perturbations,
and it is shown that the intrinsic degrees of freedom
are well behaved in the infra-red part
of the spectrum. 
We conclude that quantum fluctuations  
do not change the dynamics of the
spacetime in a way which makes  
its evolution non-perturbative 
and stochastic.  
This result contradicts previous  
derivations which are based
on the study of a quantum field on
a fixed geometry.

\end{abstract}

\section{Introduction}
It is by now widely believed that the universe
has gone through an early period of nearly exponential expansion
(see e.g. \cite{inflation} for an overview).
Apart from solving the horizon, flatness and monopole
problems, the assumption of nearly exponential expansion
has the important implication of producing a nearly scale
independent spectrum of perturbations, which is 
in agreement with observations \cite{smoot}.
Quantum fluctuations of the field which drives inflation, the inflaton, 
are generally believed to be the source of the perturbations
which gave rise to the observed structure
in the universe at late times.
An idea which goes further has been proposed by Linde 
\cite{linde} - \cite{linde4},
who argues that quantum fluctuations of the inflaton
grow in amplitude and decohere during inflation.  
Due to decoherence, the state of the inflaton can 
be described by a probability distribution, and
due to quantum fluctuations, the time evolution 
of the inflaton becomes stochastic. 
Although the mean field of the inflaton tends to
evolve to a minimum of its potential, the spread 
of the probability distribution in time allows for
a motion away from the minimum of the potential,
with some finite probability.
If there is a sufficiently large probability of finding the inflaton at a 
larger value of the potential, then inflation could proceed 
indefinitely in certain regions of the spacetime, which
then dominate in terms of the spatial volume due
to the expansion, 
while inflation may have come to an 
end in other regions.
The large-scale structure of the spacetime can then 
be expected to be very irregular, hence the name
`stochastic inflation'.
In this paper we will reconsider Linde's idea of stochastic 
inflation, by studying the evolution of a quantized
scalar field which drives inflation on a {\em perturbed}
rather than a fixed geometry.
The important difference between our derivation and
the derivations in \cite{linde} - \cite{linde4} 
arises from the definition and the interpretation
of the perturbations.
Since in previous treatments one considered inflaton 
perturbations on a fixed geometry, it was possible
to define a probability for finding the inflaton field in 
a specific state at some time. 
This picture breaks down when one considers quantum 
fluctuations of the
inflaton field, 
which are coupled
to the geometry, 
since then there is no notion
of a point in space and time at which one can compare
perturbations of the inflaton field.  
Instead, one has to deal with the intrinsic perturbations
of the inflaton and the geometry, on a background
geometry which can be taken to be a
Friedmann-Lemaitre-Robertson-Walker
(FLRW) universe.
If the intrinsic perturbations of the inflaton and the
geometry tend to become nonperturbatively large
at late times or large distances, then this shows
that the spacetime evolves away from FLRW, as
is the case in the stochastic inflation picture.
We will show that this does not happen
when the coupling between the inflaton
and the geometry perturbations is taken
into account.


The paper is structured as follows.
In section \ref{s11} we discuss the generation of long-range
quantum correlations of a scalar field, in an unperturbed 
exponentially expanding FLRW spacetime.  
It appears that these quantum correlations diverge
in the limit where the mass of the field tends
to zero.  This effect can be explained from two different
viewpoints; in comoving coordinates we find that 
individual quantum fluctuations `freeze out', while 
in a static coordinatization the quantum field 
appears to `heat up'.
 
In section \ref{C&Q} we discuss the issue of decoherence,
and the generation of classical perturbations from
quantum perturbations.

The quantization of
the inflaton and geometry perturbations is considered in section
\ref{s44}, where
we apply  Bardeen's approach \cite{bardeen} 
to describe the intrinsic degrees of freedom of the
system.
The spectrum of intrinsic quantum perturbations
is obtained in the idealized case of vanishing 
inflaton mass 
and exact exponential 
expansion,
and we study the nature 
of the geometry perturbations at 
late times and
large length 
scales.

It is shown that the intrinsic deviation of the perturbed
spacetime from FLRW does not grow at large length 
scales or at late times.
This observation, and the observational bound
on the amplitude of the perturbations,
justify our perturbative approach
to the problem.
We conclude in section \ref{C}
with the observation that no 
nonperturbative effects
occur of the kind 
which could drive 
stochastic inflation.

From now on we choose our units so that
$\hbar = c = 8 \pi G = 1$, and we let
Greek indices run from 0 to 3,
while Latin indices run from
1 to 3. 

\section{The generation of fluctuations}\label{s11}

In this section we will consider the generation of 
quantum fluctuations in an exponentially expanding
FLRW spacetime, and its analytic extension which is
de Sitter spacetime.
The use of exactly exponentially expanding FLRW
has the advantage that there are exact results
available (see e.g. 
\cite{motola}
\cite{bunch&davies}),
which is not the case for more general 
inflating spacetimes.  
Although de Sitter spacetime and a nearly exponentially
expanding
FLRW spacetime 
are topologically very
different, there are good reasons to believe that locally 
defined physical quantities are approximately the same
in these two spacetimes.
This observation is known as the `no hair conjecture' 
for exponentially expanding spacetimes \cite{gibbons&hawking}. 
The metric of spatially flat exponentially expanding FLRW
in terms of comoving coordinates reads,
\be\label{fds}
\ \mb{d} s^2 = (H\eta )^{-2}
\ (- \mb{d} \eta^2 + \delta_{i j}  \mb{d} {\bf x}^i  \mb{d} {\bf x}^j ),
\ee
where $\eta \in {\bf R}^-$ denotes the conformal time parameter, 
and $H$ is the Hubble constant.
Let us consider a free Klein-Gordon field $\phi$ with a mass
$m$ in the geometry (\ref{fds}).
In the case where the quantum state of the $\phi$-field is invariant
under spatial translations, the field operator  
which describes the quantized $\phi$-field
can be expanded in terms of a basis of modes which  
are the product of normalized plane waves $Q$,
which we define in the appendix, 
and a time dependent factor $\phi_0$ which solves the equation
\be\label{feqpsi0}
\ \ddot{\phi_0} - 2 \eta^{-1} \dot{\phi_0} 
\ +  ({\bf k}^2 + (\fr{m}{H \eta})^2 ) \phi_0 =0 ,
\ee
where a dot denotes differentiation with respect to the
conformal time $\eta$, and ${\bf k}$ denotes the
comoving wavenumber.
In the case where the mass $m$ of the field is zero, the
solutions of equation (\ref{feqpsi0}) are
given by,
\be\label{solpsi0}
\ \phi_0 = c  \eta^{\fr{3}{2}} H^{(1)}_{\fr{3}{2}} (-\bf{k}\eta),
\ee
where $H^{(1)}_{q}$ denotes the Hankel function of the first kind, 
and $c$ is a normalization constant \cite{hankel}.
At late times, i.e. when $\eta  \uparrow 0$,
the massless solutions  (\ref{solpsi0}) approach a nonzero constant,
while massive solutions of equation (\ref{feqpsi0}) go to
zero as $(-\eta)^{\fr{m^2}{3H^2}}$.
Note that although the amplitude of the modes remains 
large at late times when 
$ m \ll  H$, the
probability density 
$j_0 = \fr{H \eta}{2 {\it i}} (\phi^* \dot{\phi} - \dot{\phi^*} \phi )$
which is associated with the
modes   
decreases as fast as $\eta^3$.
The dynamical picture of the quantum fluctuations in an exponentially expanding
FLRW spacetime is that their {\em amplitude} remains large at late
times, in spite of exponentially vanishing
probability densities.
In terms of the quantization which is natural in comoving 
coordinates, there is no change in the occupation 
number of different modes, and for this reason
there is no creation of particles \cite{fulling}.


Let $G_0 ({\bf x})$ denote the two point correlation function
of the massless $\phi$-field which is assumed to be in
the vacuum state which is invariant under spatial 
translations. The correlation function $G_0 ({\bf x})$ is 
given by a sum over
modes, which factorize in terms of the time dependent solutions
$\phi_0$,  
and the spatially dependent  
harmonics $Q$
(see e.g. \cite{fulling}),
\be\label{G0X}
\ G_0 ({\bf x}) =
\ \fr{1}{2}  \int \mb{d}^3 {\bf k}  | \phi_0 (\eta, {\bf k} ) |^2 
\ Q^* ({\bf k},{\bf x}) Q ({\bf k},0). 
\ee
Let $G_0 ({\bf k})$ denote the spatial Fourier transform of
$G_0 ({\bf x})$, defined by expression (\ref{fk})
in the appendix. From expression (\ref{G0X}) it follows
directly that
\be\label{G0K}
\ G_0 ({\bf k}) = \fr{1}{2} | \phi_0 (\eta , {\bf k}) |^2.
\ee
Using expression (\ref{solpsi0}) for $\phi_0$, and the
asymptotic behavior of the Hankel function, 
$H^{(1)}_q (z) \sim z^{-q}$ for $z \downarrow 0$,
we find the asymptotic expression
for $G_0 ({\bf k})$,
\be\label{G0KforKto0}
\ G_0 ({\bf k}) \sim {\bf k}^{-3},
\ee
for ${\bf k} \downarrow 0$.
Expression (\ref{G0KforKto0}) cannot be integrated
about ${\bf k} =0$, and therefore the
correlation function $G_0 ({\bf x})$, which is
formally defined by expression (\ref{G0K}),
diverges for arbitrary ${\bf x}$.
This is the well known infra-red divergence of the
correlation function for a massless scalar field
in an exponentially expanding FLRW spacetime
\cite{vilenkin} - \cite{allen}.

If one is interested in the time dependence of 
quantum correlations,
then equation 
(\ref{G0KforKto0}) is somewhat difficult to interpret,
since the comoving wavenumber ${\bf k}$,
and the physical wavenumber
$k$, are related by the  
time dependent expression 
$k = H \eta {\bf k}$. 
In order to describe the time dependence 
of quantum correlations, it is natural
to consider the 
Fourier transform of the 
correlation 
function (\ref{G0X}) with respect
to harmonics which are normalized
in {\em physical} $k$-space.
From expression (\ref{G0K}) and the relation 
(\ref{ff}) in the appendix,   
it follows that the Fourier
transform of $G_0 ({\bf x})$
in physical $k$-space is
given by 
\be\label{GKGK}
\ G_0 (k) = (H \eta)^{-3} G_0 ({\bf k}).
\ee
The asymptotic behavior of
$G_0 (k)$  
follows from 
expression (\ref{G0KforKto0}) and (\ref{GKGK}), i.e.,
\be\label{hhh}
\ G_0 (k) \sim k^{-3},
\ee
for $k \downarrow 0$ and $\eta$ constant, and there is no time
dependence in this equation.

The infra-red divergence in equation (\ref{hhh}) 
is related to the global properties of the
exponentially expanding FLRW 
geometry.
If one assumes
that the exponential
expansion has started 
at a finite time in the past,
then one can show that $G_0 ( k)$ is well 
behaved about $k =0$ \cite{vilenkin}.
This follows e.g. by calculating the infra-red contribution 
to the expectation value of 
$|\phi|^2$,
\be\label{phisquare} 
\  \langle | \phi |^2 \rangle =
\ \int_0^{\Lambda} \mb{d}^3  k 
\  G ( k),
\ee
where $\Lambda$ is an upper boundary on the
$k$-space integral.
It can be shown that  
$\langle | \phi  |^2 \rangle$ grows
linearly in time 
when the exponential
expansion  
has started at a finite
time in the past,
and $\langle |\phi |^2 \rangle$ approaches
a constant for a positive mass $m$
\cite{vilenkin}.

A different perspective from which one can look at the 
quantum correlations in an exponentially expanding FLRW
spacetime,
is the presence of a thermal distribution of particles 
which is natural with respect to a static coordinate 
system. 
On the analytic extension of an exponentially expanding FLRW
spacetime, which
is de Sitter spacetime, one can choose coordinates
so that the metric on a section of the spacetime 
takes the static form 
\be\label{sds}
\ \mb{d} s^2 = H^{-2}  ( -  \sin^2  \theta   \mb{d} t^2 +  \mb{d} \theta^2  
\ + \cos^2 \theta \mb{d} \Omega^2 ),
\ee
where $\theta \in [ 0, \pi/2 ]$ is related to the commonly used 
radial coordinate $r$ by 
$r = H^{-1} \cos \theta $,
and $\mb{d} \Omega^2$ is the surface element
on the unit two-sphere.
We may now define quanta of the $\phi$-field, for which
the wave function $\phi_{\omega l m}$ factorizes in terms 
of a time dependent plane wave $ (2 \omega)^{-\fr{1}{2}} e^{-{\it i} \omega t}$, 
a two-sphere harmonic $Y_{lm}$, and a function of $\theta$ 
and $\omega, l,m$.
Gibbons and Hawking \cite{gibbons&hawking} showed that when
the quantum state of the field $\phi$ is analytic at the
horizon $\theta =0$, then quanta with wave functions 
$\phi_{\omega l m}$ are present with nonzero 
occupation numbers,
\be\label{df}
\ n_{\omega l m} = ( e^{ 2 \pi \omega / H} - 1)^{-1}.
\ee
In a spatially flat FLRW spacetime, a thermal 
distribution of the form (\ref{df}) does not 
give rise to divergent long range correlations, even for $m=0$,
since in this case there is a phase space element of the 
form $\mb{d}^3 k$ which cancels the divergence 
in expression (\ref{df}) at $\omega =0$.
In the quantization which is based on a static
coordinatization of de Sitter spacetime,
the situation is however different
since a constant 
time section in the metric (\ref{sds}) is one half of a
three-sphere.
The phase space in which the distribution function 
(\ref{df}) is evaluated consists therefore of
the continuous variable $\omega$, and the discrete
labels $l,m$.  

Let us now consider the massless case where $m=0$,
and $\omega \in {\bf R}^+$.
The infra-red contribution to the vacuum 
expectation value of $|\phi|^2$, is obtained 
by taking the integral over $\omega$ about
$\omega \approx  0$, of the occupation number, 
multiplied by the squared modulus 
of the wave function. 
Indeed, the occupation number $n_{\omega 00}$ 
and the square of $\phi_{\omega 00}$ diverge  
as $ \omega^{-1}$ near $\omega =0$,
and their formal integral over $\omega$ 
diverges. 
Recall that in the quantization which is based on 
comoving coordinates we found an infra-red
divergent vacuum expectation value of $|\phi|^2$,
which is due to the `freezing' of zero point fluctuations,
while no particles are created. 
In the quantization which is based on static coordinatization, the vacuum 
expectation value of $|\phi|^2$
is also infra-red divergent, but this time it is due to the 'heating' 
of the quantum field on spatially bounded sections.

\section{Classical and quantum perturbations}\label{C&Q}

In the previous section, we discussed the generation of
quantum correlations in an exponentially expanding FLRW
spacetime.
In the quantization which is based on comoving coordinates, we 
found quantum fluctuations with large amplitudes,
although in this description the 
occupation number remains zero 
for each mode 
\cite{fulling}.
The assumption that large amplitudes imply large 
occupation numbers seems to play a role in a number
of references \cite{linde} - \cite{mijic},
where it is used to argue that long range correlations
become classical in some sense.

The question how quantum correlations 
at early times evolve into classical 
perturbations at late times, 
appears to receive little attention 
in the literature,
although efforts towards a more 
satisfactory description have been made 
(see e.g. \cite{calzetta&hu} - \cite{boy}).
Without going into much detail, let us go through some of  
the ideas which play a role in the transition from quantum
to classical
perturbations.
What appears to be crucial, is
the {\em decoherence} of 
inequivalent histories \cite{coleman}.
In the context of the discussion which is given in the previous section, 
a history $\phi (x)$ is a configuration of the $\phi$-field 
which is given over the entire spacetime.
In physically relevant situations, it is often the case that
only some degrees of freedom of the $\phi$-field are 
accessible to observation, and it is natural to consider 
the equivalence classes of histories which cannot be 
distinguished by observation. 
While the state of the $\phi$-field is generally given by a
density matrix, a `reduced density matrix' can be obtained
by projecting the density matrix onto sets of inequivalent
histories.
Decoherence occurs when the reduced density
matrix is approximately diagonal, and in this
case the diagonal elements can be interpreted
as approximate probabilities for inequivalent
histories.
When two or more sets of histories are mutually exclusive at
a classical level (e.g., as in Schr\"{o}dinger's experiment
where there are two states of a cat in a box),
then they must have decohered to a high degree of
accuracy, in order to account for the absence
of observed interference effects.


Note that there is a considerable amount of freedom to choose 
a criterion by which histories are
considered to be equivalent.
Which criterion is natural in a certain situation, depends
on the type of measurement which is conducted.

As an example, let us consider again the case of a free scalar 
field $\phi$, which is in the translation invariant 
vacuum state in an exponentially expanding FLRW
spacetime.
A geodesic observer in this spacetime perceives an event
horizon at the Hubble distance $H^{-1}$, and can only respond
to quanta which originate from somewhere within
this event horizon. 
One can show that a geodesic observer responds
to quanta which have a positive frequency with
respect to the observers proper time (see e.g.
\cite{gibbons&hawking}).
These
quanta are naturally 
defined in terms of a
quantization which is 
based on a 
static coordinate system of the form (\ref{sds}), 
where the 
coordinates are chosen so that
the center of the coordinate 
system at $\theta= \fr{\pi}{2}$  
coincides with the worldline
of the geodesic observer.
When expressed in terms of the quantization which is based on static
coordinates,
the state of the $\phi$-field is described by a
reduced density matrix, which appears to be
completely diagonal \cite{gibbons&hawking}.
Decoherence is in this case obtained by summing over
the degrees of freedom of the $\phi$-field which are outside the 
event horizon.
However, the notion of an event horizon 
is observer dependent, and the reduced 
density matrix is in this case only natural 
to describe the measurement of one 
specific geodesic observer.

In the case of the early universe, there are no observers 
which can absorb quanta of a scalar field, and one
has to consider what one could call a 
`measurement situation' 
\cite{unruh&zurek} \cite{coleman}.
Essentially this means that one introduces a coupling of
the $\phi$-field with itself or other fields 
(for instance by adding a cubic interaction term 
in the Lagrangian, or by considering the
nonlinear coupling between the $\phi$-field and gravitational 
degrees of freedom).
The effect of an interaction term is that quantum fluctuations
at large length scales (compared to the Hubble radius $H^{-1}$), 
are now coupled to quantum fluctuations at small 
length scales.
By summing over perturbations of the $\phi$-field with a
wavelength which is small compared to the Hubble radius,
one obtains a reduced density matrix which describes
the decohered large scale fluctuations of the $\phi$-field
\cite{calzetta&hu}-\cite{boy}.

Let us now assume that decoherence occurs, so that
one can define a probability distribution 
$P (\phi,t)$ which describes the probability $P$ of finding
the quantum field with a value $\phi$ at a point in the spacetime with time
$t$.
In Linde's approach to stochastic inflation \cite{linde}-\cite{linde4}, 
the time evolution
of the probability distribution $P (\phi,t)$ is described
by a diffusion equation,
\be\label{FP}
\ \fr{\partial P(\phi,t)}{\partial t} =
\  D \fr{\partial^2 P (\phi,t)}{\partial^2 \phi},
\ee
where $D$ is a positive constant.
The spread of the quantum field $\phi$ about its mean value
$\langle \phi \rangle$, is given by,
\be\label{last}
\ \langle |\phi^2| \rangle = \int \mb{d} \phi |\phi - \langle \phi \rangle |^2 
\ P (\phi, t).
\ee
Recall that in section \ref{s11} we calculated $\langle |\phi^2| \rangle$
for the case of a free field in exponentially expanding FLRW.
Although the expression was found to be divergent in the
case of exact exponential expansion, a linear growth in
time can be found when the exponential expansion
has started at a finite time in the past \cite{motola}.
This result can be used to determine the value of the
diffusion constant $D$
in expression (\ref{FP}) (see e.g. \cite{linde3}).

An important limitation of the derivation which we
sketched above, is that the probability distribution
$P (\phi,t)$ is defined for an FLRW geometry. 
However, since we are interested in perturbations at 
length scales large compared to the Hubble radius,
there is no good reason why one can ignore
the coupling between the inflaton perturbations 
and the geometry \cite{bardeen}.
But if one does consider the coupling between the
inflaton perturbations and the geometry, then 
we do not have a single geometry.
Instead, for every configuration of the inflaton, one
has a different geometry, and there is no notion of a single
point in space and time for which one can define
a probability $P(\phi, t)$ for finding the field 
with a value $\phi$ at time $t$.
Indeed, one can show that quantum fluctuations of  
the inflaton field can be interpreted equally well  
as quantum fluctuations 
of only the metric, in a coordinate system 
where there is no inflaton perturbation
at all.


In the following section, we derive  
the spectrum of {\em intrinsic} 
perturbations of the inflaton and
the geometry which are generated 
during inflation. 

\section{Quantum fluctuations and stochastic inflation}\label{s44}
 
In the previous sections, we studied the amplification
of quantum fluctuations of a scalar field in an
exponentially expanding FLRW spacetime.
The correlation function was found to be infra-red
divergent in the case of a massless scalar field,
while a finite result 
was found for a positive scalar field mass 
or when we assumed that the exponential expansion 
had started at 
a finite time 
in the past. 
In the physically more realistic case of a
universe where approximate exponential
expansion is driven by a scalar field with a 
potential term, 
the scalar field mass
will generally be nonzero and time  
dependent, and the expansion
will tend to slow down gradually
before inflation ends. 
However,
since our aim is to investigate the possibility 
of an unbounded growth of 
intrinsic perturbations
during
inflation, we will neglect 
the mass of the inflaton
and the slowing down of the
expansion while the inflaton
moves down the potential.   
In our perturbative approach to describe 
the evolution of quantum fluctuations 
of the inflaton 
and the geometry during inflation, we
will therefore adopt a background geometry of 
the form (\ref{fds}). 
 As is usual in cosmology, we define a  
perturbation 
of a physical quantity as the difference of the same  
physical quantity, evaluated at corresponding points 
in a perturbed and 
a background spacetime.
What one calls a perturbation therefore depends on a mapping
between points in a perturbed and a background spacetime,
which is 
called a gauge.
In the following we will use Bardeen's formalism  
\cite{bardeen} to deal with the degrees of 
freedom which, to linear order, do not depend on the 
choice of gauge.
In our description of a scalar field in a perturbed
spatially flat FLRW spacetime, it will be sufficient to 
consider only scalar perturbations, and the
metric can be expanded as

\[
\ g_{\mu\nu} = S^2 \sum_{{\bf k}} 
\ \left[ -\delta^0_{\mu} \delta^0_{\nu} (1+ 2 A) Q \right.
\ + (\delta^0_{\mu} \delta^i_{\nu} + \delta^i_{\mu} \delta^0_{\nu}) B Q_{;i}
\  + h_{\mu\nu} (1+2 H_L Q)
\]
\be
\  \left.  + 2 H_{T } h^i_{\mu} h^j_{\nu}
\  ({\bf k}^{-2} Q_{; i ; j}+ \fr{1}{3}\delta_{ij} Q) \right],
\ee
where $;i$ denotes the derivative with respect to the
comoving spatial coordinate ${\bf x}^i$, 
$h^{\mu}_{\nu} := \delta^{\mu}_{\nu} - \delta^{\mu}_0\delta^0_{\nu}$
is the projection operator onto spatial hypersurfaces
of constant time in the background, 
and the harmonics $Q=Q ({\bf x},{\bf k})$ are
solutions of the scalar Helmholtz equation 
which are normalized according to equation
(\ref{norm}) in the appendix.  
We assume that inflation
is driven by a scalar field,
which we call $\psi$ in order  
to make a distinction with 
the field $\phi$ which was 
defined on a fixed geometry 
in the previous 
sections. 
The Lagrangian 
density of the field $\psi$ is given by 
\be\label{L}
\ {\cal L}_{\psi} = \fr{1}{8\pi G} R - 
\ g^{\mu\nu} \partial_{\mu} \psi \partial_{\nu} \psi
\ - V (\psi),
\ee
where the potential term $V (\psi)$ 
is bounded from below.
It will be useful to write the field
$\psi$
as the sum of a perturbation
$\delta \psi ({\bf x},t)$,
and the mean value
$\psi_0 (t)$,
which is defined as the average of $\psi$,
evaluated on a spatial
hypersurface 
of constant time 
in the background.

The mass of the field $\psi$ is defined as the curvature
of the potential, i.e.
$m^2 := \fr{1}{2} \partial^2 V (\psi) / \partial \psi^2$,
which is evaluated at $\psi = \psi_0$.
Nearly exponential expansion
occurs when the
kinetic term in the
Lagrangian (\ref{L}) is
small compared to the potential
term (so that $T^{\mu}_{\nu}
\approx \fr{1}{2} \delta^{\mu}_{\nu} V (\psi_0)$),
and $V(\psi_0)$
is approximately
constant on a
timescale
of the
expansion
time.
It can be shown that nearly exponential
expansion can occur for a large class
of potentials, 
including potentials of the polynomial form
$V (\psi) = \lambda \psi^{2n}$,
with $\lambda \in {R^+}$ and 
$n \in {\bf Z^+}$ 
(see e.g. \cite{linde}).

In the following, we will denote perturbations by their
Fourier components, which we defined in the appendix, 
and we
drop the argument ${\bf k}$, unless there is a
risk of confusion.

A gauge invariant variable which is useful in the description of
perturbations is given by,
\be\label{phim}
\ \phi_m :=  \delta \psi  + \fr{\dot{\psi}_0}{H}  
\ ( H_L + \fr{1}{3} H_T ),
\ee
and $\phi_m$ can be interpreted in 
terms of the three-curvature of 
the constant-$\psi$ hypersurfaces,
\be\label{R3}
\ {\cal{R}}_{\mb{constant} \psi}  =  \fr{4H}{\dot{\psi}_0} 
\ \fr{{\bf k}^{2}}{S^2} \phi_m ,
\ee
where we assume that $\dot{\psi_0} \neq 0$.

Copying Bardeen's notation, we define $\epsilon_m$ as the 
fractional energy density 
perturbation in the comoving time-orthogonal 
gauge, i.e.,
\be\label{epsilon}
\  \epsilon_m := \delta T^{0}_0 / T^0_0 ,
\ee
evaluated in the gauge where $\delta \psi = B =0$.

A variable related to $\epsilon_m$ is $\alpha_m$, which
is defined as the fractional lapse function perturbation,
evaluated in the comoving time-orthogonal gauge, i.e.
$\alpha_m := A$ in the gauge where $ \delta \psi = B =0$.
From the expression for the energy density of the scalar field
in the comoving time-orthogonal gauge, i.e.
$T^0_0 = \fr{1}{2} ( g^{00}  \dot{\psi}^2 +  V (\psi))$,
it follows that
\be\label{ea}
\ \epsilon_m = - ( 1 + \omega ) \alpha_m    ,
\ee
where $\omega :=-1 +  2 / ( 1 + V (\psi) / \dot{\psi_0}^2 )$.
The fractional energy density perturbation $\epsilon_m$  
is coupled to the entropy perturbation 
$\eta_e$, defined as the difference between the fractional
isotropic pressure in the perturbed spacetime, and 
the fractional isotropic pressure at a point in the 
background spacetime with the same energy density.
The entropy perturbation $\eta_e$, and the fractional 
lapse function perturbation $\alpha_m$, can be shown to be
proportional
\cite{bardeen},
\be\label{qui}
\ \alpha_m = \fr{\omega}{(1 + \omega) (c_s^2 -1)} \eta_e,
\ee
where we specialized Bardeen's equation (5.20) to the
case with scalar field matter, and the speed of sound
$c_s^2$ can be expressed in terms of background 
variables.
Hence, $\epsilon_m$, $\alpha_m$ and $\eta_e$ all represent the
same physical perturbation, but unlike $\epsilon_m$ and 
$\alpha_m$, the interpretation of $\eta_e$
is not related to the properties of some collection 
of spatial hypersurfaces in the perturbed spacetime, and
it is therefore an intuitively clear measure of an intrinsic
perturbation.
%
%


The gravitational and inflaton action has been expanded to
second order in terms of the scaled perturbation
variable $w := S \phi_m $ and background quantities,
by Deruelle et al. \cite{deru1}.
Recall that the variable $\phi_m$,
defined by expression (\ref{phim}), is
only to linear order 
invariant under a
gauge transformation.
The square of a first order gauge invariant
variable, is however gauge invariant to
second order.
An expansion of the action which contains
terms of linear and quadratic 
order in $\phi_m$, 
is not in general gauge
invariant to second order,
unless the linear
terms in the expansion
vanish. 
Note that an expansion of the action in terms of the
perturbation variable $\phi_m$, is obtained by integrating
the expansion of the
Lagrangian (\ref{L}) and the volume element,
over the background 
spacetime.
Indeed, terms which are linear in 
$\phi_m$ vanish in this integration,  
due to the orthogonality
relation for the harmonics $Q$
(equation (\ref{norm}) in the appendix).

The field equation for $w$ appears to be of a simple
Klein-Gordon type with a time dependent mass term,
\be\label{ky}
\  (\fr{{\partial}^2}{\partial \eta^2 } +  
\ {\bf k}^2 - 2 \eta^{-2} ) w({\bf k}) = 0,
\ee
which has the solutions  
\be\label{w}
\ w ({\bf k}) = - \fr{ \sqrt{-\pi \eta}}{4} H^{(1)}_{\fr{3}{2}} 
\ ( - \eta {\bf k}  ),
\ee
where $H^{(1)}_q$ denotes the Hankel function of the
first kind \cite{hankel}.
Note that equation (\ref{ky}) is derived
from an expansion of the action 
of the inflaton field
{\em and} the 
geometry, in terms of the gauge
invariant variable $w$ \cite{deru1}. 
The coupling between the
inflaton and the
geometry is therefore
accounted 
for
by the requirement that the solutions (\ref{w})
extremize the combined
gravitational and 
inflaton action. 

The field operator which is associated with the second quantized 
variable $w$ can
be expanded in terms of the solutions (\ref{w}), and the
two-point correlation function for the field
$w$ takes the usual form 
\cite{fulling},   
\be\label{G0KW}
\ G_0 ({\bf k}) = \fr{1}{2}
\ |w ({\bf k}) |^2,
\ee
where we assumed that the field is in the vacuum
state which is natural in comoving coordinates.
In the following we assume that the quantum state of the $w$-field
has decohered (see section \ref{C&Q}), in which case 
expression (\ref{G0KW}) 
can be interpreted
as the sum over decohered configurations 
$\{w\}$ of the square of the perturbation 
component $w ({\bf k})$ evaluated for $\{ w\}$, 
multiplied by the probability 
of finding the
configuration $\{ w \}$.

The asymptotic behavior of $G_0 ({\bf k})$ near ${\bf k}=0$
follows from expression (\ref{w}), (\ref{G0KW})
and the 
expansion of the Hankel
function, $H^{(1)}_q (z) \sim z^{-q}$ for $z \downarrow 0$. We find 
\be\label{GOKWA}
\ G_0 ({\bf k}) \sim \eta^{-2} {\bf k}^{-3},  
\ee
for ${\bf k} \downarrow 0$ and $\eta$ constant.
Assuming decoherence, expression (\ref{GOKWA})  
describes the asymptotic
behavior of the {\em classical}
perturbation component $w({\bf k})$
for ${\bf k} \downarrow 0$, 
\be\label{vk}
\  w({\bf k})  \sim \eta^{-1} {\bf k}^{-\fr{3}{2}},
\ee
where here and in the following we neglect a 
probabilistic factor of order one on the
right-hand side of the $\sim$ symbol.

\subsection{Energy and lapse function perturbations}\label{s44sub1}
Let us now determine the spectrum of fractional 
energy density and lapse function perturbations.
The fractional energy density perturbation $\epsilon_m$ has been
evaluated in terms of $w$ in \cite{deru2},
\be\label{et-tu}
\ T^0_0 \epsilon_m = 2 \fr{\kappa H}{S^2} [ \dot{w} -  H S w],
\ee
where $\kappa$ is a constant which depends on the
background geometry.
Using expression (\ref{w}) for $w$, and the differentiation property
of the Hankel function,
$( \mb{d} / \mb{d} z ) H_q (z) + (q/z) H_q (z) = H_{q -1} (z)$,
we find,
\be
\ T^0_0 \epsilon_m =\fr{\kappa \sqrt{\pi}}{S^2}
\  {\bf k} \sqrt{- \eta}
\ H^{(1)}_{\fr{1}{2}} (- {\bf{k}} \eta ),
\ee
where we used that the scale factor $S$ equals $(H\eta)^{-1}$ in the case 
of exponential expansion.
Using the asymptotic expression for the Hankel
function, $H^{(1)}_q (z) \sim z^{-q}$ for $z \downarrow 0$, yields,
\be\label{Tf}
\  \epsilon_m \sim \alpha_m \sim \eta^2 {\bf k}^{\fr{1}{2}}.
\ee
The time dependence in equation (\ref{Tf})  
is a consequence of the  
Fourier decomposition of the perturbations 
in comoving rather than physical $k$-space.
Let $\bar{\epsilon}_m$ and $\bar{\alpha}_m$
denote the physical $k$-space 
Fourier components of
the
fractional
energy density and lapse 
function perturbations,
which we define by expression 
(\ref{fk-phys}) in the appendix.
According to expression (\ref{ff}) in the appendix, the perturbation
components in physical
and comoving $k$-space 
are related by $\bar{\epsilon}_m = S^{\fr{3}{2}} {\epsilon}_m$
and $\bar{\alpha}_m = S^{\fr{3}{2}} \alpha_m$. 
Using this relation, and expression (\ref{Tf}),
yields the asymptotic expression,
\be\label{Tfp}
\  \bar{\epsilon}_m \sim \bar{\alpha}_m
 \sim  {k}^{\fr{1}{2}},
\ee
where $k = S^{-1} {\bf k}$ denotes
the physical wavenumber, and there
is no time dependence at the
right-hand side of equation
(\ref{Tfp}). 
Expression (\ref{Tfp}) shows that the infra-red
behavior of the fractional density
perturbation and lapse function
is regular,
and
integrable 
about 
$k =0$.

Note that expression (\ref{Tfp})  
determines the asymptotic behavior of the spectrum of 
fractional energy density 
perturbations, but not its magnitude.
Observations of
perturbations
at the time of last
scattering \cite{smoot},
put an upper limit on
${\epsilon}_m$ of the order
of $10^{-5}$ for
wavenumbers
small compared to
the value of the Hubble parameter 
at that time.
Since $\epsilon_m$
grows approximately
as the square of the
scale factor for the
growing mode and small
wavenumbers \cite{bardeen},
the fractional energy density
perturbations
must have been many
orders of magnitude
smaller than
$10^{-5}$
at the time when
inflation came
to an end.

\subsection{Curvature perturbations}\label{s44sub2}

In this subsection, we investigate the spectrum
of curvature perturbations of the hypersurfaces
on which the inflaton field $\psi$ is constant. 
According to expression (\ref{ff}) in the appendix, 
the perturbation components in physical
and comoving $k$-space
are related by $\bar{\phi}_m = S^{\fr{3}{2}} {\phi}_m$.
Using expression (\ref{vk}), we find the asymptotic
behavior of $\bar{\phi}_m$ for $ k \downarrow 0$, 
\be\label{phimk}
\ \bar{\phi}_m  = S^{\fr{3}{2}} \phi_m \sim k^{-\fr{3}{2}} ,
\ee  
and there is no time
dependence in equation (\ref{phimk}). 

From expressions (\ref{R3}) and (\ref{vk}) 
we derive the expression for the Fourier
component of the
spatial curvature perturbation, with
respect to the physical wavenumber $k$,  
\be\label{R33}
\   \bar{\cal R}_{\mb{constant} \psi}  \sim { k}^{\fr{1}{2}},
\ee
for ${k} \downarrow 0$.

We will now address the question whether the geometry
of the constant-$\psi$ hypersurfaces
develops increasingly strong
inhomogeneities at large length scales or at late times.
More generally, one would like to know whether our perturbed
FLRW spacetime remains everywhere close to FLRW in
some well defined sense.

The question whether an inhomogeneous spacetime is close
to FLRW in some objective sense, is known as the averaging
problem in cosmology. This problem appears to be surprisingly
nontrivial, even in the linearized case, since it involves 
comparing tensors in different spacetimes, in a way which 
is independent of the choice of coordinates, gauge, and the 
averaging prescription (see e.g. \cite{jelle} and references therein).

In the following we will sidestep the difficulties involved
with a precise formulation of the averaging problem,
by considering the scaling behavior of the spectrum of
perturbations.
Without presenting an objective criterion for a spacetime
being close to an FLRW spacetime, we determine whether intrinsic
perturbations grow or decrease under a change
of the length scale.
This allows us to compare the warping of our universe
at superhorizon scales and at length scales where
observations can be made, e.g., on the intersection of
our past light cone with the surface of
last scattering.
If perturbations tend to grow at large length scales, then our
perturbative treatment can be expected to break down
at superhorizon scales, which is in agreement with the picture of a
stochastically inflating universe which is strongly warped at
superhorizon scales \cite{linde}- \cite{linde4}.

On the contrary, a decreasing or constant spectrum of perturbations
implies that our perturbative approach holds on superhorizon
scales, given that it holds at subhorizon scales.
This will then give us an improved estimate of the 
effect of quantum fluctuations on the global structure
of the universe.

We define a new length scale $\bar{\ell}$
by choosing a new unit of physical length which
is $a \in {\bf R}$ times the former unit of length.
In terms of the new length scale, a physical quantity $q$ with
dimension $\ell^d$ transforms to $\bar{q} = a^{-d} q$.
As we derived in the appendix, the Fourier transform
$q (k)$ of a quantity
$q (x)$ 
transforms under scale transformations
as a half-density of this quantity
in $k$-space, i.e.
$\bar{q}(\bar{k }) = a^{-d -\fr{3}{2}} q (k)$,
where  
$\bar{k} = a k$.

The scaled spatial curvature perturbation $\bar{\cal R} (k)$ therefore
takes the form
\be\label{R4}
\ \bar{\cal{R}}_{\mb{constant} \phi} = 
\  a^{\fr{1}{2}} {\cal{R}}_{\mb{constant}\phi},
\ee
where we used that the spatial curvature has the
dimension $\ell^{-2}$.

Using expression (\ref{R33}) for the spectrum of three-curvature
perturbations, and the scale transformation property (\ref{R4}),
we find,
\be\label{R5}
\ \bar{\cal{R}}_{\mb{constant} \phi} \sim a^{\fr{1}{2}} k^{\fr{1}{2}}
\ =  \bar{k}^{\fr{1}{2}}.
\ee
Comparing equations (\ref{R33}) and (\ref{R5}), we 
find that the spectrum of curvature perturbations of the
constant-$\psi$ hypersurfaces remains constant
under a change of
the length scale.


Summarizing the results derived in this 
subsection, we found that perturbations
in the spatial curvature 
are regular about $k =0$,
and the perturbations do not grow
at late times or at large 
length scales.  

\subsection{Interpretation}

Let us now address the question why the
gauge invariant perturbations
$\bar{\cal R}$ and $\bar{\alpha}_m$
do not show
the same infra-red divergence
as we found for a quantum field
$\phi$ on
a fixed geometry.
At first sight, this might be surprising, since
expressions (\ref{G0K}) and (\ref{GOKWA}) show
that
the
gauge invariant perturbations $\phi_m$, and the perturbations $\phi$
on a fixed geometry,
diverge with the same
power when $k \downarrow 0$.
The essential difference between the
perturbations $\phi_m$ and $\phi$,
is their interpretation
in terms of
physical
quantities.
Note that a perturbation of the field $\phi$
on a fixed geometry
is proportional to a
perturbation in the
energy density,
which follows
by expanding the
energy density
$T^0_0 = \fr{1}{2} ( \p^0 \phi \p_0 \phi +  V (\phi))$
about the background value $\phi =\phi_0$,
where we use that $ V^{\prime} (\phi)  |_{\phi_0}  \neq 0$
as long as the scalar field has not reached a
minimum of the potential.
On the contrary, it follows from equation (\ref{Tfp}) and
(\ref{phimk}) that the gauge
invariant amplitude $\phi_m$ acts as a
{\em potential} for the fractional
lapse function and energy
density perturbations
in the limit when $k \downarrow 0$.
Similarly, expression (\ref{R3}) shows that
$\phi_m$
acts as a potential for the
spatial curvature
perturbation
${\cal R}$.
The perturbation
$\phi_m$ may therefore diverge
as $k^{-\fr{3}{2}}$ for $k \downarrow 0$,
while the spatial curvature 
${\cal R}$, and the 
fractional lapse function and energy
density
perturbations
$\alpha_m$ and $\epsilon_m$ 
are well behaved 
when $k \downarrow 0$.

One may question whether the 
time evolution
of the perturbations in the spatial curvature
and the fractional lapse function
can be described by a diffusion 
equation  
of the form 
(\ref{FP}). 
For a massless quantum field $\phi$ on a 
fixed geometry, this method 
is natural since 
the expectation 
value of $|\phi|^2$, 
which according
to equation (\ref{last}) equals to the
standard deviation of 
the probability
distribution $P (\phi,t)$, 
appeared to grow linearly
in time. 
In this section we showed that the spectrum 
of fractional lapse function
and spatial curvature 
perturbations is 
integrable
about $k=0$,
and constant 
in time.
The probability distribution which
describes the decohered fractional
lapse function and spatial curvature
perturbations, 
is therefore time independent, 
and it
does not evolve according to the 
diffusion
equation (\ref{FP}).

\section{Conclusions}\label{C} 
                             
In the previous section we 
derived expressions
for the spectrum of the 
fractional lapse function
and spatial curvature 
perturbations which are
generated during 
exponential
expansion.
We found that intrinsic perturbations 
of the scalar field and the geometry
do not grow at late times or at
large length scales, during 
exponential expansion.
The constancy of the spectrum of perturbations
in space and time, and the 
observational bounds 
on the
magnitude  
of the perturbations, 
justify our perturbative
approach.
Our result 
contradicts the assumption that 
quantum fluctuations grow 
nonperturbatively
during inflation, 
which underlies
the idea of 
stochastic 
inflation.
This 
indicates that an inflating universe is not likely
to develop a highly
irregular structure at superhorizon scales,
unless these irregularities are present
in the initial conditions.  

\section{Acknowledgements}

I would like to thank George F. R. Ellis
for helpful comments.
Thanks also go to Henk van Elst for reading the manuscript.
The research was supported with funds from
FRD (South Africa).

\section{Appendix}
 
In this appendix, we discuss the scaling behavior
of the flat space Fourier transform.
Let us define the solutions $Q ({\bf k},{\bf x})$,
which satisfy the scalar Helmholtz equation, i.e.
$Q^{;i}_{;i} = |{\bf k}|^2 Q$, where $;i$ denotes
the covariant derivative with respect to
${\bf x}^i$, and the solutions $Q$ are
normalized according to
\be\label{norm}
\ \int \mb{d}^3 {\bf x} Q^* ({\bf k},{\bf x}) Q ({\bf k}^{\pr}, {\bf x} )
\ = \delta^3 ({\bf k} - {\bf k}^{\pr}).
\ee
The Helmholtz equation and the normalization
condition (\ref{norm}) are satisfied by distributions of
the form
$Q ({\bf k},{\bf x}) = N  e^{{\it i} {\bf k}{\bf x}}$,
where the normalization factor $N$ is
defined as the distribution which is
constant for all ${\bf x}$, and which is normalized
as,
\be\label{n00}
\int \mb{d}^3 {\bf x} N ^2  =(2\pi)^{-3}.
\ee
The Fourier transform of a function $f ({\bf x})$
with respect to the comoving
wavenumber ${\bf k}$
is given
by,
\be\label{fk}
\ f ({\bf k}) = \int \mb{d}^3 {\bf x} Q ({\bf k},{\bf x}) f ({\bf x}).
\ee
Let us now consider how
$f ({\bf k})$ transforms under
scale transformations.
We implement a scale transformation by defining
a new unit of length $\bar{\ell}$,
which equals $a \in {\bf R}$ times
the former
unit of length.
In terms of the new length scale, we define solutions $\bar{Q}$ 
of the Helmholtz equation,
which
are normalized
by the condition
\be\label{norm2}
\ \int \mb{d}^3 \bar{x} \bar{Q}^*(\bar{k},\bar{x}) \bar{Q}(\bar{k}^{\pr},\bar{x})
\ = \delta^3 (\bar{k} - \bar{k}^{\pr}).
\ee
A basis of solutions $\bar{Q}$ which satisfy the
normalization condition (\ref{norm}) is
given by
$ \bar{Q} (k,x) = \bar{N} e^{{\it i}\bar{k}\bar{x}}$,
where $\bar{N}$
is defined as the constant
distribution
which satisfies the normalization condition
\be\label{n11}
\ \int \mb{d}^3 \bar{x} \bar{N} ^2  =(2\pi)^{-3}.
\ee
It follows from the definitions (\ref{n00}) and (\ref{n11})
that the normalization constants $N$ and $\bar{N}$
are related
by $\bar{N} = a^{\fr{3}{2}} N$, which
implies
\be\label{QQ}
\  \bar{Q} (\bar{k},\bar{x}) = a^{\fr{3}{2}} Q (k,x),
\ee
where $\bar{x} = a^{-1} {\bf x}$ and $\bar{k} = a {\bf k}$.
The Fourier transform of a function with respect to
the barred wavenumber $\bar{k}$ is defined by
\be\label{fk-phys}
\ {f} ( \bar{k}) := \int \mb{d}^3 \bar{ x}
\ \bar{ Q} (\bar{k},\bar{x}) f (\bar{x}),
\ee
and from the relation (\ref{QQ})
and the definitions
(\ref{fk}) and (\ref{fk-phys}) it follows that
\be\label{ff}
\ f (\bar{k}) = a^{-\fr{3}{2}} f ({\bf k}),
\ee
where $\bar{k}= a{\bf k}$.
Expression (\ref{ff}) shows that the Fourier transform
of a function $f (x)$ transforms under a scale
transformation as a half-density in $k$-space.
In some calculations it will be useful to express
the Fourier transform with respect
to the comoving wavenumber (\ref{fk}),
in terms of the Fourier transform with
respect to the physical wavenumber.
Since the comoving wavenumber and the
physical wavenumber
are related by a time dependent
scale transformation $k = S^{-1} {\bf k}$,
this is simply a special case of the
scale transformation discussed above,
and
$f (k)$ and $f ({\bf k})$
are related
by $f (k) = S^{\fr{3}{2}} f ({\bf k})$,
where $k = S^{-1} {\bf k}$.


\begin{thebibliography}{30}
\bibitem{inflation} K. A. Olive, Phys. Lett. {\bf 190}, No. 6, 307 - 413
(1990).
\bibitem{smoot} G. F. Smooth et al., ApJ., 396, L1 (1992).
\bibitem{linde} A. Linde, in Particle Physics and Inflationary
Cosmology, Harwood Academic Publishers (1990).

\bibitem{linde3} A. Linde, Int. J. Mod. Phys. {\bf A}, Vol. 2, No. 3,
561 (1987).
\bibitem{linde4} A. Linde, Phys. Rev. {\bf D 49}, 1783 (1994).
\bibitem{mijic} M. Mijic, Phys. Rev. {\bf D 57}, 2198 (1998).
\bibitem{bardeen} J. M. Bardeen, Phys. rev. {\bf D 22}, 1882 (1982).
\bibitem{vilenkin} A. Vilenkin, Nucl. Phys. {\bf B 226}, 527 (1983).
\bibitem{motola} E. Motola, Phys. Rev. {\bf D 31}, 754 (1984).
\bibitem{bunch&davies} T. S. Bunch and P. C. Davies, Proc. R. Soc.
Lond. A. {\bf 360}, 117-134 (1978).
\bibitem{allen} B. Allen, Phys. Rev. {\bf D 32}, 3136 (1985).
\bibitem{fulling} S. A. Fulling, {\em Aspects of Quantum
Field Theory in Curved Space-Time}, Cambridge University Press (1989).
\bibitem{gibbons&hawking} G. W. Gibbons and S. W. Hawking,
Phys. Rev. {\bf D 15}, 2738 (1977).
\bibitem{unruh&zurek} W. G. Unruh and W. H. Zurek,
Phys. Rev. {\bf D 40}, 1071 (1988).
\bibitem{calzetta&hu} E. Calzetta and B. L. Hu, Phys. Rev. {\bf D 52}, 6770 (1995).
\bibitem{calzetta&hu2} E. Calzetta and B. L. Hu, Phys. Rev. {\bf D 49}, 6636 (1993).
\bibitem{esteban} A. Esteban et al., Phys. Rev. {\bf D 55}, 1812 (1997). 
\bibitem{hu&matacz} B. L. Hu and A. Matacz, Phys. Rev. {\bf D 49}, 6612 (1994).
\bibitem{boy} D. Boyanovski, Phys. Rev. {\bf D 57}, 2166 (1998).
\bibitem{coleman} J. B. Hartle, in {\em Quantum cosmology
and baby universes},  edited by S. Coleman {\it et. al.}, World
Scientific (1991).
\bibitem{jelle} J. Boersma, Phys. Rev. {\bf D 57}, 1890 (1998).
\bibitem{deru1} N. Deruelle et al., Phys. Rev. {\bf D 45},
3301 (1992).
\bibitem{deru2} N. Deruelle et al., Phys. Rev. {\bf D 46},
5337 (1992).
\bibitem{hankel} I. S. Gradhsteyn, I. M. Ryshik, {\em Tables of Integrals,
Series and Products} (Academic, New York, 1980).
\end{thebibliography}
\end{document}